\documentclass{iaus}
\usepackage{graphicx}

\newcommand{\n}{{\nabla}} 
\newcommand{\kv}{{\bf k}}

\newcommand{\ta}{\Theta} 
\renewcommand{\vec}[1]{{\bf #1}}

\title[Dynamos of Giant Planets] 
{Dynamos of Giant Planets}

\author[Busse \& Simitev]   
{F. H. Busse$^1$%
   \and R. Simitev$^2$}
\affiliation{$^1$Institute of Physics, University of Bayreuth, D95440 Bayreuth, Germany
 \break email: busse@uni-bayreuth.de\\[\affilskip]
$^2$Department of Mathematics, University of Glasgow, Glasgow G12 8QW, 
 UK \break email: rs@maths.gla.ac.uk}

\pubyear{2006}
\volume{239}  
\pagerange{467--474}
\date{?? and in revised form ??}
\jname{Convection in Astrophysics}
\editors{F. Kupka, I.W. Roxburgh \& K.L. Chan, eds.}
\begin{document}

\maketitle

\begin{abstract}
Possibilities and difficulties of applying the theory of magnetic field generation by convection flows in rotating spherical fluid shells to the Giant Planets are outlined. Recent progress in the understanding of the distribution of electrical conductivity in the Giant Planets suggests that the dynamo process occurs predominantly in regions of semiconductivity. In contrast to the geodynamo the magnetic field generation in the Giant Planets is thus characterized by strong radial conductivity variations. The importance of the constraint on the Ohmic dissipation provided by the planetary luminosity is emphasized. Planetary dynamos are likely to be of an oscillatory type, although these oscillations may not be evident from the exterior of the planets.
\end{abstract}

\section{Introduction}

Even before the Pioneer and Voyager missions evidence for a global
Jovian magnetic field had been obtained through the observations of
decametric radio waves by \cite{BF}. The subsequent discoveries of the
magnetic fields of Saturn, Uranus and Neptune have been highlights of
the exploration of the solar system by space probes. While Jupiter and
Saturn possess magnetic fields which are dominated by a dipole part
that is nearly aligned with the axis of planetary rotation, Uranus and
Neptune exhibit magnetic fields that do not show such an alignment and
are also characterized by a strong quadrupolar component. This latter
property has been interpreted in earlier papers (\cite{RS}; \cite{SB})
as a result of the thin shell to which the dynamo process is assumed
to be confined. This explanation is not entirely convincing, however,
since \cite{LGS} argue convincingly that the dynamo process in Jupiter
and Saturn is also confined to thin fluid shells. As will be pointed
out in section 5 magnetic fields without a dominant axis-aligned
dipole are a rather common feature of numerical simulations of
planetary dynamos. 

 In the following we shall first briefly introduce the mathematical
 background for numerical dynamo simulations. At the present stage it
 appears to be appropriate to introduce only a minimum of physical
 parameters in order to obtain an overview of the parameter dependence
 of planetary dynamos. Hence we shall not even consider the
 compressibility of the fluid in the Giant Planets and restrict the
 analysis to the Boussinesq approximation. An alternative approach has
 been used by \cite{EG} (see also article by Glatzmaier in these
 Proceedings) where a fairly accurate representation of the
 compressibility of Jupiter has been attempted.

\section{Mathematical Description of Convection Driven Spherical Dynamos}

For the equations describing convection driven
dynamos in the frame rotating with the angular velocity $\Omega$ 
we use a standard formulation which has also been used for a dynamo
benchmark (\cite{C}). A more general  
static state will be assumed with the temperature distribution $T_S = T_0 - \beta
d^2 r^2 /2 + \Delta T \eta r^{-1} (1-\eta)^{-2}$ where $\eta$ denotes
the ratio of inner to outer radius of the spherical shell and $d$ is its thickness.
$\Delta T$ is the temperature difference between the boundaries in the
special case $\beta =0$. In the case $\Delta T = 0$ the static temperature distribution $T_S$ corresponds to that of a homogeneously heated sphere with the heat source density proportional to the parameter $\beta$.
The gravity field is given by $\vec g = - \gamma d \vec r$ where
$\vec r$ is the position vector with respect to the center of the sphere and
$r$ is its length measured in units of $d$. 

In addition to  $d$, the
time $d^2 / \nu$,  the temperature $\nu^2 / \gamma \alpha d^4$ and 
the magnetic flux density $\nu ( \mu \varrho )^{1/2} /d$ are used as
scales for the dimensionless description of the problem  where $\nu$ denotes
the kinematic viscosity of the fluid, $\kappa$ its thermal diffusivity, $\varrho$ its density and $\mu$ its magnetic permeability.
The Boussinesq approximation is used in that $\varrho$ is assumed to be
constant except in the gravity term where its temperature dependence given by
$\alpha \equiv - ( d \varrho/dT)/\varrho =$ const. is taken into
account. The dimensionless equations of motion, the heat equation for
the deviation $\Theta$ of the temperature field from the static
distribution and the equation of magnetic induction thus assume the
form 
\begin{equation}
\n^2 {\bf v} + \vec B \cdot \nabla \vec B + \vec r \ta - \n \pi \nonumber  = P^{-1}(\partial_t\vec v + \vec v\cdot\n \vec v) 
+ \tau \kv \times \vec v 
\end{equation}
\begin{equation}
 \n \cdot \vec v = 0
\end{equation}
 \begin{equation}
 \nabla^2 \Theta + \left[ R_i +R_e \eta r^{-3} (1 - \eta)^{-2} \right]\vec r \cdot \vec v \nonumber = P ( \partial_t + \vec v \cdot \nabla ) \Theta
\end{equation}
\begin{equation}
P_m(\partial_ t +\vec v \cdot \nabla)\vec B + \nabla \times (\frac{\lambda}{\lambda_0} \nabla \times \vec B)  = P_m\vec B \cdot\nabla \vec v
\end{equation} 
where $\kv$ is the unit vector in the direction of the axis of rotation and where $\n \pi$ includes all terms that can be written as gradients. The Rayleigh numbers $R_i$ and $R_e$, the Coriolis parameter $\tau$, the Prandtl
number $P$ and the magnetic Prandtl number $P_m$ are defined by
\begin{equation}
 R_i = \frac{\alpha \gamma \beta d^6}{\nu \kappa} , 
\enspace R_e = \frac{\alpha \gamma \Delta T d^4}{\nu \kappa} ,
\enspace \tau = \frac{2
\Omega d^2}{\nu} ,\enspace P = \frac{\nu}{\kappa} , \enspace P_m = \frac{\nu}{\lambda_0}
\end{equation}
where $\lambda_0$ is a typical value of the magnetic diffusivity $\lambda$ which we allow to vary as a function of the distance $r$ from the center in contrast to the other material properties. Such a variation reflects the often significant variation of the electrical conductivity $\sigma$ with radius according to the relationship $\lambda = \sigma^{-1}\mu^{-1}$. While $P=1$ is often assumed with the argument that all
effective diffusivities are equal in turbulent media we keep $P$ as a parameter. 

Since the velocity field $\vec v$ as well as the magnetic flux density 
$\vec B$ are solenoidal vector fields,  
the general representation in terms of poloidal and toroidal
components can be used,
\begin{equation}
\vec v = \nabla \times ( \nabla \Phi \times \vec r) + \nabla \Psi \times 
\vec r \enspace ,
\enspace \vec B = \nabla \times  ( \nabla h \times \vec r) + \nabla g \times 
\vec r \enspace .
\end{equation} 
By multiplying the (curl)$^2$ and the curl of equation (6a)
by $\vec r$ we can obtain two equations for $\Phi$ and $\Psi$ which will not be given here explicitly (see, for example, \cite{SB5}).
The equations for $h$ and $g$ are obtained through the multiplication of
equation (2.4) and of its curl by $\vec r$,
\begin{equation}
[2(1-a)(r - r_i) + a]\nabla^2 L_2 h = P_m [ \partial_t L_2 h - \vec r \cdot \nabla \times ( \vec v
\times \vec B )] 
\end{equation}
\begin{equation}
[2(1-a)(r - r_i) + a]\nabla^2 L_2 g + 2(1-a)r^{-1}\partial_r(rg) = P_m [ \partial_t L_2 g - \vec r \cdot \nabla \times ( \nabla
\times ( \vec v \times \vec B ))]
\end{equation}
where the $L_2$ is defined by 
\begin{displaymath}
L_2 \equiv - r^2 \nabla^2 + \partial_r ( r^2 \partial_r)
\end{displaymath}
In (2.7) and (2.8) a linear dependence $\lambda = \lambda_0[2(1-a)(r - r_i) + a]$ has been introduced such that $\lambda = \lambda_0$ at the middle of the layer, $r = r_i + 0.5$. 

Either rigid boundaries with fixed temperatures as in the benchmark case (\cite{C}),
\begin{equation}
\Phi = \partial_{r}(r\Phi) = \Psi = \Theta = 0 \enspace \mbox{ at }
\enspace r=r_i \equiv \eta / (1- \eta) 
\enspace \mbox{ and at } \enspace r=r_o \equiv
(1-\eta)^{-1},
\end{equation}
or stress-free boundaries
with fixed temperatures,
\begin{equation}
\Phi = \partial^2_{rr}\Phi = \partial_r (\Psi/r) = \Theta = 0 \enspace \mbox{ at }
\enspace r=r_i 
\enspace \mbox{ and at } \enspace r=r_o,
\end{equation}
are frequently used. The latter conditions allow to cover numerically a larger region of the
parameter space since the thin Ekman layers at the boundaries are nearly absent. The radius ratio $\eta = 0.4$ is often used in the simulations since it provides a good compromise for the
study of both, the regions inside and outside the tangent
cylinder. The latter is the virtual cylindrical surface touching the inner
spherical boundary at its equator. 
For the magnetic field it is convenient to employ electrically insulating
boundaries such that the toroidal component of the field vanishes there, while the poloidal function $h$ must be 
matched to the function $h^{(e)}$ which describes the  
potential fields 
outside the fluid shell,  
\begin{equation}
g = h-h^{(e)} = \partial_r ( h-h^{(e)})=0 \; 
\mbox{ at } r=r_i\enspace \mbox{ and at } \enspace r=r_o .
\end{equation}
Alternatively an infinitely conducting inner boundary can be assumed,
\begin{equation}
\partial_r g = h=0 \; 
\mbox{ at } r=r_i,
\end{equation}
but computations for the case of an inner boundary with no-slip
conditions and an electrical conductivity equal to that of the fluid
are also often done.
The numerical integration of equations (2.3), (2.7), (2.8), (2.9) and (2.10) together with boundary
conditions proceeds most often with the pseudo-spectral 
method as described by \cite{GL} and \cite{TI}
which is based on an expansion of all dependent variables in
spherical harmonics for the $\theta , \phi$-dependences, i.e. 
\begin{equation}
\Phi = \sum \limits_{l,m} V_l^m (r,t) P_l^m ( \cos \theta ) \exp \{ im \phi \}
\end{equation}
and analogous expressions for the other variables, $\Psi, \Theta, h$ and $g$. 
$P_l^m$ denotes the associated Legendre functions.
For the $r$-dependence expansions in Chebychev polynomials are used. 

It should be emphasized that the static state $\vec v = \vec B = \Theta = 0$ represents a solution of equations (6) for all values of the Rayleigh numbers $R_i$ and $R_e$, but this solution is unstable except for low or negative values of the latter parameters. Similarly, there exist solutions with $\vec B = 0$, but $\vec v \not= 0, \Theta \not= 0$, for sufficiently large values of either $R_i$ or $R_e$ or both, but, again, these solutions are unstable for sufficiently large values of $P_m$ with respect to disturbances with $\vec B \not= 0$. Dynamo solutions, as all solutions with $\vec B \not= 0$ and $|\vec B| \propto r^{-3}$ for $r\rightarrow \infty$ are called, are thus removed by at least two bifurcations from the basic static solution of the problem. 

\section{Convection in rotating spherical shells}

Past research on convection driven dynamos in rotating, self-gravitating spherical shells has shown that the properties of convection in the absence of a magnetic field do change quantitatively, but not qualitatively after the Lorentz force enters the force balance. It is thus important to understand the properties of non-magnetic convection. A rough idea of the dependence of the critical Rayleigh number $R_{c}$
for the onset of convection on
the parameters of the problem can be gained
from the expressions derived from the annulus model (see recent review of \cite{BU})
\begin{subeqnarray}
\hspace{0cm} R_{c} = 3 \left( \frac{P \tau }{1+P} \right)^{\frac{4}{3}} ( \tan
\theta_m)^{\frac{8}{3}} r_m^{-\frac{1}{3}} 2^{-\frac{2}{3}}, \hspace{6cm} 
\\ \hspace{-0.5cm}m_c = \left( \frac{P \tau}{1+P} \right)^{\frac{1}{3}} ( r_m \tan
\theta_m )^{\frac{2}{3}} 2^{-\frac{1}{6}} ,\quad
\omega_c = \left( \frac{\tau^2}{(1+P)^2P} \right)^{\frac{1}{3}}
2^{-\frac{5}{6}} 
(\tan^2 \theta_m / r_m )^{\frac{2}{3}},
\end{subeqnarray}
where $r_m$ refers to the mean radius of the fluid shell, $r_m = (r_i + r_o)/2$,
and $\theta_m$ to the corresponding colatitude, $\theta_m =$ arcsin $(r_m(1-\eta))$.
The azimuthal wavenumber of the preferred mode is denoted by $m_c$ and the
corresponding angular velocity of the drift of the convection columns in the
prograde direction is given by $\omega_c / m_c$. 

\begin{figure}[ht]
\begin{center}
\includegraphics[angle=-180,width=6cm]{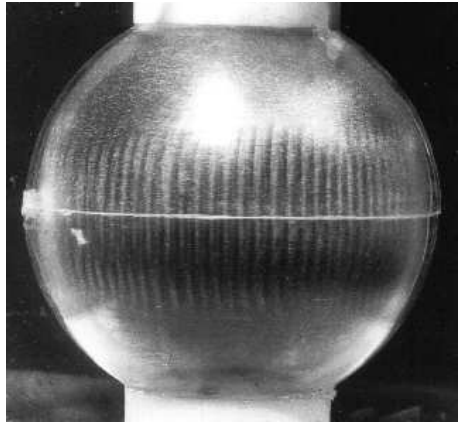}
\end{center}
\caption{Banana cells in a thin rotating spherical fluid shell cooled
from within. Convection driven by centrifugal buoyancy is made visible
by a suspension of tiny flakes which become aligned with the
shear. Because the inner sphere is cooled, the outer heated, the experiment models the planetary situation where gravity and temperature gradient are reversed since only the product enters force balance. (After \cite{BC})} 
\label{f3}
\end{figure}

While expressions (3.1) correspond to motions in the form of columns aligned with the axis of rotation, convection in the form of ``banana cells" is realized in less rapidly rotating or thinner spherical shells as experimentally visualized in figure 1. An analytical theory of the ``banana cells" which includes   the differential rotation generated by their Reynolds stresses was first derived by \cite{B7}. At low Prandtl numbers, namely for $P\lesssim 10/\sqrt{\tau}$ inertial convection in the form of equatorially attached cells becomes prevalent (\cite{A}; \cite{SB3}). Its name reflects the fact that it can be described as a small modification of certain inertial waves (\cite{Z4}; \cite{BS}) 
A third form of convection is realized in the polar regions of the
shell which are defined as the two fluid domains inside the tangent
cylinder. Since gravity and rotation vectors are nearly parallel in
these regions (unless $\eta\equiv r_i/r_o$ assumes a value close to unity) convection
resembles the kind realized in a horizontal layer heated from below
and rotating about a
vertical axis. Because the Coriolis force can not largely be balanced by the pressure gradient in this case, the onset of convection is delayed to higher values of $R$ where convection outside the tangent cylinder has reached already high amplitudes. 

\begin{figure}[t]
\begin{center}
\hspace*{0cm}
\includegraphics[height=2.5in]{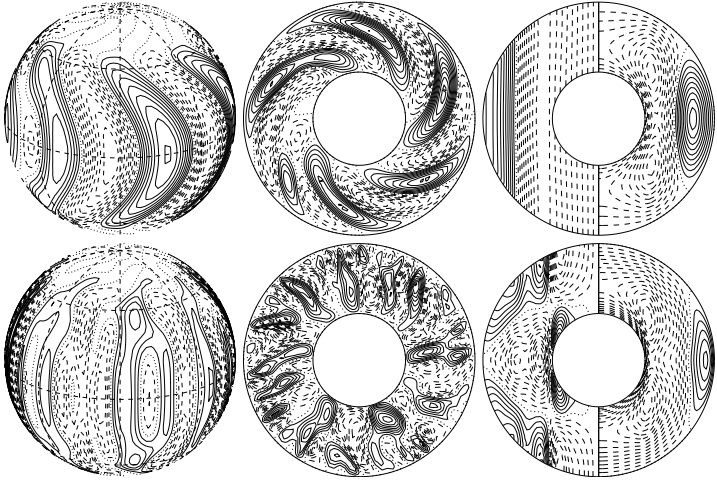}
\end{center}
\caption{Convection in rotating spherical fluid shells for
  $\tau = 5\cdot10^3$, $R=2.7 \cdot R_c$, in the cases $P=0.5$ (upper row, $R= 1.5 \cdot 10^5$) and  $P=20$  (lower row, $R= 4 \cdot 10^5$). Lines of
  constant $u_r$ in the middle spherical surface, $r=r_i +0.5$,
  are shown on the right. The middle plots show
  streamlines, $r \partial \Phi / \partial \varphi=$ const., in the
  equatorial plane. The plots on the right indicate lines of constant
   mean azimuthal velocity $\bar u_{\varphi}$ in the left halves
  and isotherms of $\bar{\Theta}$   in the right halves.  }
\label{f2}
\end{figure}
More important than its influence on the onset of convection according to relationships (3.1) is the effect of the Prandtl number on convection at finite amplitudes. Typical features of low and high Prandtl number convection are
illustrated in figure 2. The columnar nature of convection does not
vary much with $P$ as is evident from the two plots on the left side of the
figure. At Prandtl numbers of the order unity or less, - but not in
the case of inertial convection-, the convection columns tend to
spiral away from the axis and thereby create a Reynolds stress which
drives a strong geostrophic differential rotation as shown on the
right side of the figure. This differential rotation in turn
promotes the spiral tilt and a feedback loop is thus created. At high values
of $P$ the Reynolds stress becomes negligible and no significant mean tilt
of the convection columns is apparent in the middle plot of the lower row. In this case the differential
rotation is generated as a thermal wind caused by the
latitudinal gradient of the axisymmetric component of $\Theta$. 

\begin{figure}[t]
\begin{center}
\hspace*{0cm}
\includegraphics[height=2.6in]{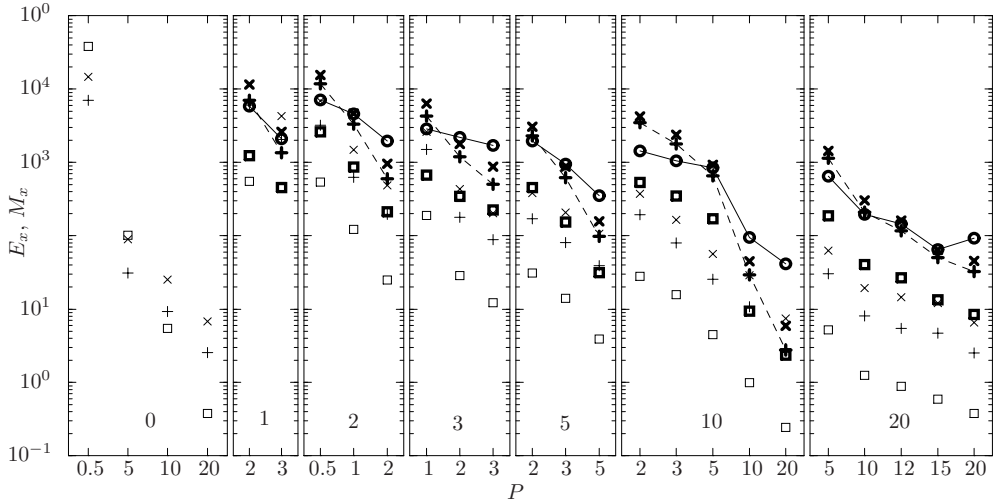}
\end{center}
\caption{Kinetic (thin symbols) and magnetic (thick symbols) energy densities of convection driven dynamos as function of $P$ for
  $\tau = 3\cdot10^4$, $R=3.5\cdot 10^6$, $\eta=0.4$, $a=1$ (implying $\lambda \equiv \lambda_0$) and values of $P_m$ as indicated in the boxes. The components $\overline{X}_p,  \overline{X}_t, \check{X}_p \check{X}_t$ (where $X = E$ or $M$) are represented by circles, squares, plus-signs and crosses, respectively.}
\label{f3}
\end{figure}

The properties of convection are reflected in the averages of the kinetic energy densities of the various
components of the velocity field which are defined by
\begin{subeqnarray}
\overline{E}_p = \frac{1}{2} \langle \mid \nabla \times ( \nabla \bar \Phi \times \vec r )
\mid^2 \rangle , \quad \overline{E}_t = \frac{1}{2} \langle \mid \nabla \bar \Psi \times
\vec r \mid^2 \rangle, \\
\check{E}_p = \frac{1}{2} \langle \mid \nabla \times ( \nabla \check \Phi \times \vec r )
\mid^2 \rangle , \quad \check{E}_t = \frac{1}{2} \langle \mid \nabla \check \Psi \times
\vec r \mid^2 \rangle,
\end{subeqnarray}
where the angular brackets indicate the average over the fluid shell
and $\bar \Phi$ refers to the azimuthally averaged component of $\Phi$,
while $\check \Phi$ is defined by $\check \Phi = \Phi - \bar \Phi$. Analogous definitions hold for the magnetic energy densities where $E, \Phi$ etc. are replaced by $M, h$ etc., 
In figure 3 energy densities have been plotted for convection with and without magnetic fields.

\section{Convection driven dynamos}

Dynamos are generated by convection in rotating
spherical shells for all parameter values  as long as the magnetic
Reynolds number, $Rm\equiv P_m \sqrt{2E}$ is of the order $50$ or higher and the fluid is not
too turbulent where the
kinetic energy density $E$ is defined by
$E=\overline{E}_p+\overline{E}_t+\check{E}_p+\check{E}_t$. In planetary cores $P_m$ assumes values of the order of $10^{-6}$ and less, but numerical
simulation have achieved so far only values somewhat below
$10^{-1}$. 
An important feature demonstrated in figure 3 is the change in
the structure of the magnetic field with increasing Prandtl
number. While for low values of $P$ the mean poloidal field is
small in comparison with the fluctuating components, this situation
reverses as $P$ increase. This change is associated with the transition from
the geostrophic differential rotation to the thermal wind type differential rotation caused by a latitudinal
temperature gradient. While the magnetic energy $M$ may exceed the total kinetic energy $E$ by orders of magnitude in particular for high Prandtl numbers, ohmic dissipation is usually found to be roughly comparable to viscous dissipation or less in numerical simulations. This may be due to the limited numerically accessible parameter space, however.

\begin{figure}
\begin{center}
\hspace*{0cm}
\includegraphics[height=2.4in]{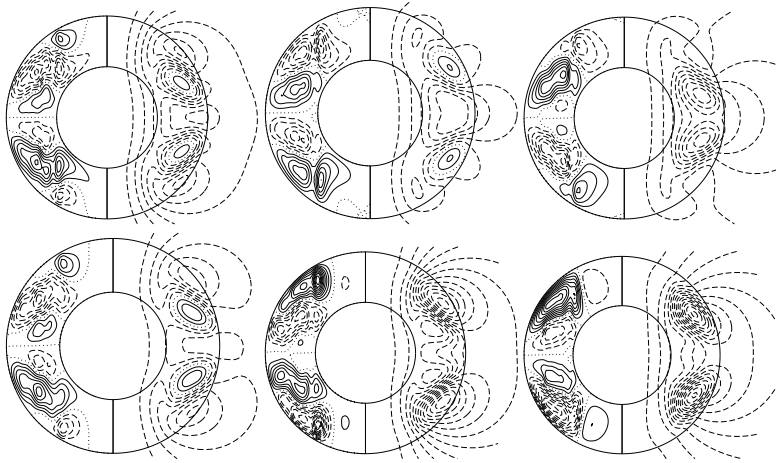}
\end{center}
\caption[]{An ``invisibly'' oscillating dynamo 
in  the case of $P=2, P_m=1$, $\tau=3 \cdot10^4, \eta=0.5, a=1$ and $R=2 \cdot 10^6$.
The plots show lines of constant
$\overline{B_{\varphi}}$
in their left halves and meridional field lines, $r \sin \theta
\partial_\theta \overline{h}  =$ const., in their right halves.
The time sequence of plots starts at the upper left and continues clockwise with $\Delta t = 0.04$  such that a full cycle is completed approximately.} 
\label{f4}
\end{figure}

As long as convection does not occur in the polar regions or is
sufficiently weak there, convection outside the tangent cylinder is
nearly symmetric with respect to the equatorial plane as is evident
from figure 2. As a consequence the dynamo generated magnetic fields
exhibit a dipolar or a quadrupolar character unless the
non-axissymmetric components by far exceed the axisymmetric components
of the field as sometimes happens for high values of $Rm$ or low
values of $P$. In addition dynamos of hemispherical character are
often found, for which the field in the northern half is much weaker
than in the southern half or vice versa. Quadrupolar and hemispherical
dynamos typically oscillate. Dipolar dynamos often oscillate as well,
but sometimes these oscillations can not be identified at a distance
of a radius or more as is evident in the example of figure 4. 

\section{Applications to the Giant Planets}

It is obvious from the preceding sections that numerical simulations
are still far removed from realistic descriptions of dynamo processes
in the Giant Planets. There are even more basic issues that need to be
resolved. Since the Proudman-Taylor-Theorem holds for barotropic
fluids the strong differential rotations observed at the surface of
the planets must be expected to continue for a considerable distance
into the planetary interiors. Because the poloidal electric current
density will be of the order $U B_p \sigma$ where $U$ is a typical
zonal velocity, say $100 m/s$, and $B_p$ measures the strength of the
poloidal field the density of Ohmic dissipation becomes of the order
$(U B_p)^2\sigma$. When this expression is integrated over a
reasonable domain inside the planet it turns out (\cite{LGS}) that it
exceeds easily the net luminosity of Jupiter as well as that of Saturn
when the value of $\sigma$ for metallic hydrogen is used. Accordingly
\cite{LGS} argue that the dynamo is located in a region where not only
the strong zonal flows penetrating from the surface have been
truncated, but where also  $\sigma$ is as small as is compatible with
a magnetic Reynolds number of the order of a few $10^2$. Such a region
would lie at a depth of less than $15\%$ $(35\%)$ of the planetary
radius of Jupiter (Saturn) where hydrogen is still a semiconductor
(\cite{N}). A consequence of the latter property is a strong
temperature dependence of $\sigma$ which translates into a high radial
increase of $\lambda$. It will be of interest to see how these
properties and constraints can be accommodated in dynamo simulations
producing magnetic fields similar to the observed ones. 

The dynamos operating in Uranus and Neptune are similarly constrained
(\cite{HB}). Traditionally an ionic electrical conductivity has been
assumed for  the dynamo regions of these planets, but recent
measurements (\cite{L}) indicate that water is semiconducting under
the relevant conditions. 
\begin{figure}
\begin{center}
\includegraphics[height=2in]{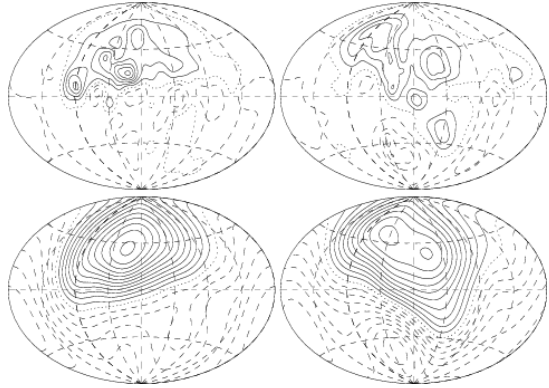}
\end{center}
\caption[]{Convection driven dynamos 
in  the cases  $P=0.3, P_m=2, R_i= 2.5\cdot 10^6$ (left) and $P=0.5, P_m= 2.5, R_i=2.2\cdot 10^6$ (right).
The values $\tau= 3\cdot 10^4$, $\eta=0.4$ and $a=0.2$ are the same in both cases. The plots show lines of constant $B_r$ on the surface
  $r=r_o+0.7$ (upper row) and $r=r_o+1.5$ (lower row)}. 
  \label{f5}
\end{figure}

We close with some remarks on the unusual form of the magnetic fields of Uranus and Neptune. The properties that dipole components of the magnetic fields are relatively weak in comparison with other components and that  the dipoles are not aligned with the axes of rotation are not especially unusual.
Magnetic fields of such character can often be found in simulations of convection driven dynamos. It seems that they are more often found when the diffusivity varies with radius. In figure 6 some typical examples are shown which seem to resemble the observed magnetic fields of the Icy Planets.

\begin{discussion}

\discuss{Andy Ingersoll}{Jupiter's field is generated so close to the surface, i.e. at $0.86 R_J$. Why isn't the field ``rougher'' at the surface?}

\discuss{F.H. Busse}{The smoothness depends also on the effective Prandtl numbers $P$ and $P_m$.}
\end{discussion}


\begin{thebibliography}{}

\bibitem[Ardes et al. 1997]{A}
{Ardes, M., Busse, F. H., \& Wicht, J.} 1997, 
 \textit {Phys. Earth Plan. Int.} 99, 55

\bibitem[Burke \& Franklin (1955)]{BF}
     {Burke, B.F. \& Franklin, K.L.} 1955,
     \textit{J.~Geophys. Res.} 60, 213

\bibitem[Busse (1970)]{B7}
{Busse, F. H.} 1970,
 \textit {ApJ} 159, 629

\bibitem[Busse, 2002]{BU}
{Busse, F. H.} 2002,
 \textit {Phys. Fluids} 14, 1301

\bibitem[Busse \& Carrigan, 1976]{BC}
{Busse, F. H., \& Carrigan, C. R.} 1976, 
 \textit {Science} 191, 81
  
\bibitem[Busse \& Simitev, 2004]{BS}
{Busse, F. H., \& Simitev, R.} 2004, 
\textit {J. Fluid Mech.} 498, 23

\bibitem[Christensen et al., 2001]{C}
{Christensen, U.R., Aubert, J., Cardin, P., Dormy, E., Gibbons, S., Glatzmaier, G.A., Grote, E., Honkura, Y., Jones, C., Kono, M., Matsushima, M., Sakuraba. A,, Takahashi, F., Tilgner, A., Wicht, J., Zhang, K.} 2001, \textit {Phys. Earth Plan. Inter.} 128, 25

\bibitem[Evonuk \& Glatzmaier (2006)]{EG}
     {Evonuk, M., \& Glatzmaier, G. A.} 1991,  
     \textit{Icarus} 181, 458
     
\bibitem[Glatzmaier (1984)]{GL}
{Glatzmaier, G. A.} 1984,
 \textit {J. Comp. Phys.} 55, 461
     
\bibitem[Holme \& Bloxham, 1996]{HB}
      {Holme, R., \& Bloxham, J.} 1996 
      \textit {J. Geophys. Res.} 101, 2177
      
\bibitem[Lee et al., 2006]{L}
     {Lee, K.K.M., Benedetti, L.R., Jeanloz, R., Celliers, P.M., Eggert, J.H., Hicks, D.G., Moon, S.J., Mackinnon, A., Collins, G.W., Henry, E., Koenig, M., \& Benuzzi-Mounaix, A.} 2006,  
     \textit{J. Chem. Phys.} 125, 014701

\bibitem[Liu et al. (2006)]{LGS}
     {Liu, J., Goldreich, P.M., \& Stevenson, D.J.} 2006,  
     \textit{Icarus} submitted

\bibitem[Nellis et al., 1996]{N}
     {Nellis, W.J., Weir, S.T., \& Mitchell, A.C.} 1996, 
     \textit {Science} 273, 396

\bibitem[Ruzmaikin \& Starchenko, 1991]{RS}
     {Ruzmaikin, A.A., \& Starchenko, S.V.} 1991,  
     \textit{Icarus} 93, 82
     
\bibitem[Simitev \& Busse, 2003]{SB3}
     {Simitev, R., \& Busse, F.H.} 2003,
     \textit{New J. Phys.}5, 97.1
     
\bibitem[Simitev \& Busse, 2005]{SB5}
     {Simitev, R., \& Busse, F.H.} 2005,
     \textit{J.~Fluid Mech.} 532, 355

\bibitem[Stanley \& Bloxham, 2004]{SB}
     {Stanley, S., \& Bloxham, J.} 2004,  
     \textit{Nature} 428, 151

 \bibitem[Tilgner (1999)]{TI}
{Tilgner, A.} 1999,
  \textit {Int. J. Numer. Meth. Fluids} 30, 713 
    
\bibitem[Zhang, 1994]{Z4}
{Zhang, K.} 1994, 
\textit {J. Fluid Mech.} 268, 211 

\end{thebibliography}
\end{document}